\documentclass[aip,twocolumn,letterpaper]{revtex4}

\usepackage{fullpage}

\usepackage{graphics}
\usepackage{epsfig}

\begin{document}
\title{Magnetic reconnection with null and X-points }
\author{Allen H. Boozer}
\affiliation{Columbia University, New York, NY  10027\\ ahb17@columbia.edu}

\begin{abstract}

Null and X-points are not themselves directly important to magnetic reconnection because distinguishable field lines do not approach them closely. Even in a collision-free plasma, magnetic field lines that approach each other on a scale $c/\omega_{pe}$ become indistinguishable during an evolution.   What is important is the different regions of space that can be explored by magnetic field lines that pass in the vicinity of null and X-points.  Traditional reconnection theories made the assumption that the reconnected magnetic flux must be dissipated or diffused by an electric field.  This assumption is false in three dimensional systems because an ideal evolution can cause magnetic field lines that cover a large volume to approach each other within the indistinguishability scale $c/\omega_{pe}$.  When the electron collision time $\tau_{ei}$ is short compared to the evolution time of the magnetic field $\tau_{ev}$, the importance of $c/\omega_{pe}$ is replaced by the resistive time scale $\tau_\eta=(\eta/\mu_0)L^2$ with $L$ the system scale.  The magnetic Reynolds number is $R_m\equiv\tau_\eta/\tau_{ev}$ is enormous in many reconnection problems of interest.  Magnetic flux diffusion implies the  current density required for reconnection to compete with evolution scales as $R_m$ while flux mixing implies the required current density to compete scales as $\ln R_m$.  

\end{abstract}

\date{\today} 
\maketitle


\section{Introduction}

Plasma forces or changing boundary conditions cause magnetic fields embedded in plasmas to evolve. The time scale of this evolution $\tau_{ev}$ is frequently very short compared to the resistive time scale $\tau_\eta=(\eta/\mu_0)L^2$ with $L$ the spatial scale of the plasma.  When $\tau_{ev}<<\tau_\eta$, the magnetic field complexity and energy generally increase until magnetic reconnection allows a simplification of the magnetic structure.   For a simplification to occur, the rate of reconnection must reach the rate of evolution, which is difficult to achieve when the magnetic Reynolds number satisfies, $R_m\equiv \tau_\eta/\tau_{ev} \rightarrow\infty$.  

Zweibel and Yamada noted in their review of magnetic reconnection \cite{Zweibel:review} that the magnetic Reynolds number is very large indeed.  It is typically of order $10^4$ to $10^8$ in moderately large laboratory plasmas, $10^8$ to $10^{14}$ in the Sun, and $10^{15}$ to $10^{21}$ in the interstellar medium of galaxies.    

When $R_m\rightarrow\infty$, the magnetic evolution is expected to be of the ideal form
\begin{equation}
\frac{\partial \vec{B}}{\partial t}=\vec{\nabla}\times(\vec{u}_\bot\times\vec{B}), \label{ideal-B}
\end{equation}
where $\vec{u}_\bot$ is the velocity of the magnetic field lines \cite{Newcomb}.  But, mathematics implies \cite{Boozer:ideal-ev} that the ideal evolution equation fails on a time scale of order   $\tau_{ev}\ln R_m$, where the evolution time is
\begin{equation}
\tau_{ev} \equiv \frac{1}{\|\vec{\nabla}\vec{u}_\bot\|},
\end{equation}
and $\|\vec{\nabla}\vec{u}_\bot\|$ is a typical value for the largest term in the $3\times3$ tensor $\vec{\nabla}\vec{u}_\bot$.  Observations typically show magnetic reconnection events are prevalent after a time scale that is between one and two orders of magnitude longer than the evolution time.  The reconnection events are characterized \cite{Liu:2017} by phenomena that occur on a time scale that is longer but comparable to the Alfv\'en times, $\tau_A =L/V_A$.

Traditional reconnection theory, which dominates the literature and the reviews \cite{Zweibel:review,Loureiro:2016}, assumes an initial state in which the current density associated with the reconnection is concentrated in a narrow channel and has a characteristic magnitude $j\approx (B/\mu_0L)R_m$.   How a natural evolution would produce a current of this intensity is generally not explained. Nevertheless, the current density must be this intense for resistive diffusion to be as fast as the evolution \cite{Schindler:1988} and Appendix \ref{sec:traditional-reconnection}.  

Ian Craig and collaborators have found analytic solutions that give a singular current density at a null of the magnetic field \cite{Craig:2014}.  But they do not explore the implications of the lines of a magnetic field embedded in an evolving collisionless plasma being indistinguishable near a null.  This  indistinguishability will be discussed and found to be of great importance.  

What traditional reconnection theory ignores is the importance of chaotic advection, which is a mathematical concept based on Lagrangian coordinates that was introduced by Hassan Aref  \cite{Aref:1984} in 1984.  Aref's paper is of foundational importance in many applications of fluid mechanics and has more than 1300 citations.

Langragian coordinates $\vec{x}(\vec{x}_0,t)$, give the Cartesian position $\vec{x}$ of points moving with a fluid at time $t$ as a function of their Cartesian position $\vec{x}_0$ at time $t=0$, Appendix \ref{sec:Lagrangian}.   A streamline of the flow is given by $\vec{x}(\vec{x}_0,t)$ as a function of time with $\vec{x}_0$ fixed.

Chaotic advection means neighboring streamlines of the flow separate exponentially in time, which causes an exponential increase in the rate of diffusive mixing.  It should be stressed that chaotic advection does not mean the flow velocity is random or has large derivatives.  The chaotic flows considered by Aref are smooth functions of position and time.  

Complete mixing, for example of chemicals, requires not only chaotic advection but also diffusion.  Aref was well aware of this; indeed his first equation includes the diffusivity.  Nevertheless, Aref used a Lagrangian analysis to study only the advective part of the advective-diffusion equation.    The full power of the Lagrangian coordinates requires that the diffusive as well advective term be represented, which was done  \cite{Tang-Boozer:1999} by Tang and Boozer in 1999 

Aref showed  \cite{Aref:1984} that when a liquid is stirred that it is essentially inevitable that the flow is chaotic when the flow depends on three spatial coordinates or even on two spatial coordinates when the flow is time dependent.  The empirical realization of importance of chaotic advection produced by stirring on mixing goes goes back much further than Aref---at least to the beginning of cooking.      Anyone with doubts should try stirring a cup containing coffee and cream without mixing them.

Eighteen years before the foundational work of Aref on chaotic advection \cite{Aref:1984}, David Stern \cite{Stern:1966} reviewed the long history of the representation of a magnetic field that is evolving ideally, Equation (\ref{ideal-B}), using Lagrangian coordinates,
\begin{equation}
\vec{B}(\vec{x},t) = \frac{1}{J} \frac{\partial \vec{x}}{\partial\vec{x}_0} \cdot \vec{B}_0(\vec{x}_0). \label{Lagrangian-ideal}
\end{equation}
The quantity $\partial \vec{x}/\partial\vec{x}_0$ is the $3\times3$ Jacobian matrix, which is discussed in Appendix \ref{sec:Lagrangian},  $J$ is its determinate or Jacobian, and $\vec{B}_0(\vec{x}_0)$ is the magnetic field expressed in the Cartesian coordinates $\vec{x}_0$ at the initial time, $t=0$.  Equation (\ref{Lagrangian-ideal}) is valid for an arbitrary component of the velocity $\vec{u}$ along the magnetic field lines.

The mathematical basis for expecting plasma motion to be of exponential importance to the rate of reconnection have been in place for decades, since the papers of Stern \cite{Stern:1966}, Aref \cite{Aref:1984}  and Tang and Boozer \cite{Tang-Boozer:1999}.  It is an interesting historical question why these classic papers have had essentially no effect on reconnection theory, Section \ref{sec:rationale}.

 The mathematically simplest way to break the ideal evolution of Equation (\ref{ideal-B}) is given by the Ohms law for a collisionless, pressureless plasma, $\vec{E}+\vec{v}\times\vec{B}=\mu_0(c/\omega_{pe})^2 \partial \vec{j}/\partial t$, which implies
\begin{equation}
\frac{\partial}{\partial t}\left(\vec{B}+\left(\frac{c}{\omega_{pe}}\right)^2\vec{\nabla}\times\left(\vec{\nabla}\times\vec{B}\right)\right)=\vec{\nabla}\times(\vec{v}\times\vec{B}). \label{collisonless ev}
\end{equation}
As shown in Appendix  \ref{c/omega-p}, the smoothing produced by $c/\omega_{pe}$ implies that two magnetic field lines that come closer than $c/\omega_{pe}\equiv \sqrt{m_e/(\mu_0n_e e^2)}$ anywhere along their trajectories will not have their separate identities preserved during an evolution when they are embedded in a collisonless plasma.   The distinction is also lost between two magnetic field lines separated by a distance less than $\delta_\eta\equiv\sqrt{(\eta/\mu_0)\tau_{ev}}\approx(c/\omega_{pe})\sqrt{\tau_{ev}/\tau_{ei}}$ in a resistive plasma that is evolving on the time scale $\tau_{ev}$ and has an electron-ion collision time $\tau_{ei}$.  Although $c/\omega_{pe}$ is small, of order centimeters in the solar corona, chaotic advection amplifies the the importance of $c/\omega_{pe}$ exponentially.  Solar mass ejections have a scale $\approx 10^9$ times larger than $c/\omega_{pe}$ with $R_m\approx10^{12}$.

The $c/\omega_{pe}$ limitation on the distinguishability of evolving magnetic field lines is of fundamental importance to all reconnection studies since the electron is the lightest charged particle and, therefore, has the least inertia of any current carrier.
 
The existence of a scale $\geq c/\omega_{pe}$ over which magnetic field lines freely change their connections suggests a change in focus of reconnection studies from places where the current density is high to regions in which magnetic field lines that are separated by $c/\omega_{pe}$ at one location are separated by distance comparable to the system scale $L$ at others.  To be of direct importance to reconnection, such regions must fill a volume, the volume undergoing reconnection.  Filling a volume means that an arbitrarily small sphere placed anywhere within the volume must contain magnetic field lines that undergo an adequate number of exponentiations in separation.

When reconnection arises from the magnetic field lines undergoing an exponential separation from the $c/\omega_{pe}$ or $\delta_\eta$ scale to the scale of the reconnection, the current density is only relevant to the extent that it is required for consistency.  As shown below and in \cite{Boozer:B-line.sep}, the required current density for that is only $j \approx (B/\mu_0L) \ln R_m$.  For $R_m\approx10^{12}$, the difference in a current density $j \propto \ln R_m$ and the value $j\propto R_m$ expected in traditional reconnection theory is approximately the difference between the cost of a cell phone and the annual size of the U.S. economy---a non-trivial difference indeed.

A heuristic argument clarifies the importance and scaling of the effects of chaotic advection on magnetic field evolution. The equation for a magnetic field line $\vec{x}(\ell)$ is $Bd\vec{x}/d\ell=\vec{B}(\vec{x})$, where $\ell$ is the distance along the line.  The separation $\vec{\delta}$ between two neighboring magnetic field lines, $\vec{x}(\ell)$ and $\vec{x}(\ell) +\vec{\delta}(\ell)$, is given by $ B d \vec{\delta}/d\ell =( \vec{\delta}\cdot\vec{\nabla})\vec{B}$ as $|\vec{\delta}|\rightarrow0$.  The magnitude of the $3\times3$ tensor $\vec{\nabla}\vec{B}$ is approximately $\mu_0 j_{||}$, which implies the separation between the two lines can go from $\delta_0$ to $w$ as a field line goes a distance $L$ when the parallel current obeys  $(\mu_0j_{||}/B)L\approx\ln(w/\delta_0)$.  Reconnection is obtained over a region of width $w$ when two lines reach a separation $w$ that  had a separation $\delta_0$ at  $\ell=0$, which was smaller than either $c/\omega_{pe}$ or $\delta_\eta$.  This gives the scaling $j_{||}\propto\ln(R_m)$ for the current density required for reconnection.  The change in the magnetic field required to produce an adequate separation is $\delta B \approx \mu_0 j_{||}/w$ so 
\begin{eqnarray}
\delta B&\approx& \ln\left(\frac{w}{\delta_0}\right) \frac{w}{L}B \nonumber\\
&=& \ln(R_m)  \frac{w}{L} B, \hspace{0.1in}\mbox{when}\hspace{0.1in} \label{delta B}\\
\frac{L}{w} &>>& \ln(R_m). \label{L/w}
\end{eqnarray}
The force per unit area $\vec{\mathcal{F}}$ that must be applied at a boundary to produce the field $\delta \vec{B}$, which can be taken to be tangential to the boundary, is given by Equation (\ref{Force}) of Appendix \ref{Bndry-Force}, $\vec{\mathcal{F}} = (B_\bot/\mu_0)\delta \vec{B}$, where $B_\bot$ is the magnetic field perpendicular to the boundary.

The force at the boundary required to drive a system to a rapidly reconnecting state is not large.  But, the reason the force is small is subtle and without care the magnitude of the required force can appear to be large.

Magnetic field line spreading is determined by magnetic flux conservation.  Any magnetic field line can be placed in the center of a flux tube and its separation from other lines can be monitored by the behavior of the cross section of the tube.   When neighboring magnetic field lines exponentiate apart, an initially circular flux tube with a radius of $c/\omega_{pe}$ becomes high distorted; its circumference increases exponentially in length as time advances.  The greatest distance transverse to the flux tube increases only diffusively in time if the flow of the field lines $\vec{u}_\bot$ has no long-range components, but can reach the scale $L$ on the time scale $(c/\omega_{pe})\ln(t/\tau_{ev})\approx L$ when $\vec{u}_\bot$ has long-range components. Small-scale turbulent-type stirring in fluids produces slower mixing than large-scale stirring.  Turbulence enhances the rate of magnetic reconnection by a mechanism related to chaotic advection, and this is discussed in a number of papers, which have been reviewed by Eyink \cite{Eyink:2015}.  But, small-scale turbulence is intrinsically a slower mechanism than chaotic advection when the chaotic advection has a scale comparable to the scale of the magnetic reconnection. 

\begin{figure}
\centerline { \includegraphics[width=2.0in]{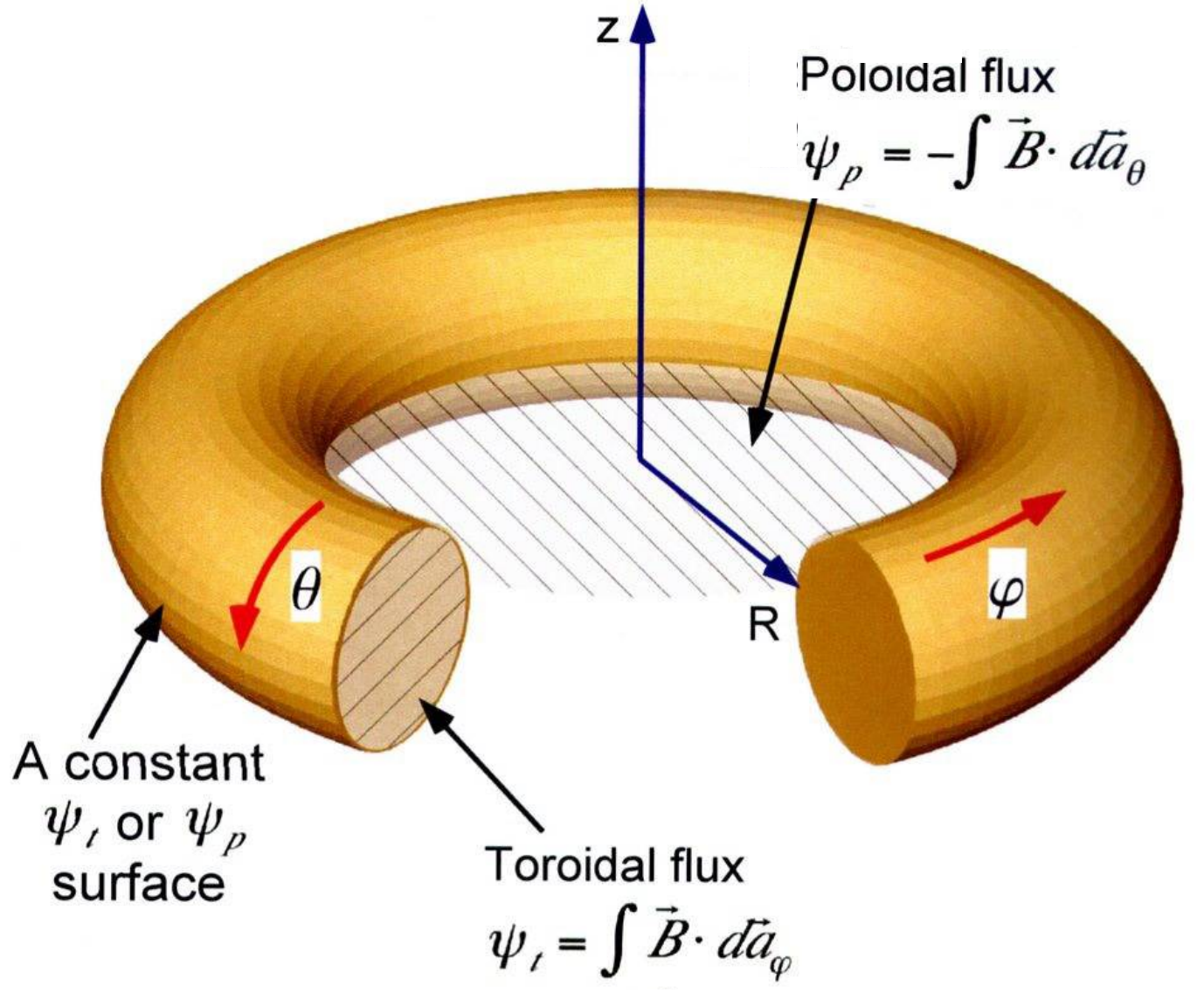} }
\caption{The toroidal flux $\psi_t$ is the magnetic flux enclosed by a toroidal surface.  The poloidal flux $\psi_p$ is the flux passing through the central hole of the torus.  Also illustrated are the poloidal $\theta$ and the toroidal $\varphi$ angles.  $(R, \varphi, Z)$ are ordinary cylindrical coordinates.   This is Figure 2 in A. H. Boozer, Nucl. Fusion \textbf{57}, 056018 (2017).}  
\label{fig:coordinate}
\end{figure}

The connection between flux and fluid stirring is particularly clear in a toroidal plasma in which a coordinate $\varphi$ exists such that $\vec{B}\cdot\vec{\nabla}\varphi=0$.  Then, an ideally evolving magnetic field has the form, Appendix to \cite{Boozer:RMP},
\begin{eqnarray}
2\pi \vec{B} =\vec{\nabla}\psi_t\times\vec{\nabla}\theta +\vec{\nabla}\varphi\times\vec{\nabla}\psi_p(\psi_t,\theta,\varphi), \label{canonical B}
\end{eqnarray}
where $\psi_t,\theta,\varphi$ and $\psi_p$ are interpreted in Figure \ref{fig:coordinate}.  The poloidal flux is the Hamiltonian for the magnetic field lines and determines their trajectories in its canonical coordinates,  $(\psi_t,\theta,\varphi)$,
\begin{eqnarray} 
\frac{d\psi_t}{d\varphi}=- \frac{\partial\psi_p}{\partial\theta} \hspace{0.2in}\mbox{and}\hspace{0.2in} \frac{d\theta}{d\varphi}=\frac{\partial\psi_p}{\partial\psi_t}. \label{magnetic Hamiltonian}
\end{eqnarray}  
A plot the magnetic field lines in ordinary Cartesian coordinates also requires the function $\vec{x}(\psi_t,\theta,\varphi,t)$.  When the $t$ dependence of $\vec{x}(\psi_t,\theta,\varphi,t)$ satisfies the mathematical conditions for a homotopy, the evolution is ideal, Appendix to \cite{Boozer:RMP}.   The velocity of the  $(\psi_t,\theta,\varphi)$ canonical coordinates through space is $\vec{u}_c=(\partial\vec{x}/\partial t)_c$.  When the velocity  $\vec{u}_c$ is both chaotic and a smooth function of position, as it is in standard chaotic advection theory based on continuous Hamiltonians, even the simple $\psi_p=\psi_p(\psi_t)$ of systems with magnetic surfaces can become an extremely complicated function of position in an ideal evolution.  A small non-ideal effect can allow a single magnetic field line to quickly change from lying on nested toroidal surfaces to coming arbitrarily close to every point in a volume.  When non-ideal effects are very small, the magnetic helicity, which is proportional to $\int \psi_p d\psi_t d\theta d\varphi$ integrated over the volume covered by the field line is the same before and after a magnetic reconnection \cite{Boozer:e-runaway2019}.  The proof of Equation (\ref{canonical B}) given in \cite{Boozer:RMP} does not depend in an essential way on toroidal geometry, and consequently has a very general validity.

Two concepts, (1) exponentiation due to chaotic advection and (2) the minimum scale $c/\omega_{pe}$ of magnetic field line distinguishability, fundamentally change not only the theory of magnetic reconnection but also the interpretation of X-points, Section \ref{sec:X-points}, and magnetic nulls, Section \ref{sec:null prop}.  These are the primary topics of this paper.  A summary is given in Section \ref{summary}.

Although the centrality chaotic advection to the mixing of fluids is documented in thousands of papers, the concept has not had significant consideration in the sixty-year effort to understand magnetic reconnection.  The focus has been and is on a near-singular current density, $j\propto R_m$.  The scientific rationales that have been given or implied for ignoring chaotic advection in reconnection studies are discussed in Section \ref{sec:rationale}.

Details are relegated to appendices to make the paper easier to read.  Appendix \ref{sec:traditional-reconnection} discusses traditional reconnection theory to allow comparisons.  Appendix \ref{sec:Lagrangian} gives the fundamental properties of Lagrangian coordinates that are required to understand this paper.  Appendix \ref{c/omega-p} derives the equation that shows that even an evolving collision-free plasma does not preserve the distinguishability of magnetic field lines on a scale smaller than $c/\omega_{pe}$.  Appendix \ref{Bndry-Force} calculates the force required to achieve a e-folding separation of magnetic field lines.  Appendix \ref{sec:stochastic example} gives a simple map that allows a study of the type of magnetic field line stochasticity that is of importance for the solar corona.  Appendix \ref{near-null traj} derives the behavior of magnetic field lines arbitrarily close to a magnetic field null.

The transfer of magnetic energy to the plasma as a whole or to individual particles when reconnection is enhanced by chaotic advection is not discussed in this paper since that was the subject of a recent publication \cite{Boozer:part.acc.}.


\section{Rationale for ignoring  Aref and Stern \label{sec:rationale} }

A number of rationales for ignoring the relevance of the work of Aref and Stern to magnetic reconnection have been given or implied.  This section shows that none of these rationales are valid.


\subsection{Force constraint}

A rationale for ignoring the implications of the work of Aref and Stern is that unlike the stirring of a cup containing coffee and cream, the stirring of a magnetic field has force implications that constrain the motions of the magnetic field lines.  Appendix \ref{sec:Lagrangian} shows that the relation  between Lagrangian coordinates and an ideal evolution, Equation (\ref{Lagrangian-ideal}), is equivalent to
\begin{eqnarray}
\vec{B}(\vec{x},t) &=& \frac{\hat{u}\cdot\vec{B}_0}{\Lambda_m\Lambda_s} \hat{U} +  \frac{\hat{m}\cdot\vec{B}_0}{\Lambda_u\Lambda_s} \hat{M} +  \frac{\hat{s}\cdot\vec{B}_0}{\Lambda_u\Lambda_m} \hat{S},  \hspace{0.2in}  \label{Exp-B}
\end{eqnarray}
when a Singular Value Decomposition of the Jacobian matrix, $\partial\vec{x}/\partial\vec{x}_0$ is carried out. The three lower case, orthonormal unit vectors are in the space of the initial Cartesian coordinates, $\vec{x}_0$, and the three upper case unit vectors are in space of the Cartesian coordinates $\vec{x}$.  The $\Lambda$'s are the three singular values.  As is standard in three dimensional chaotic advection $\Lambda_u$ increases exponentially with time, $\Lambda_s$ decreases exponentially with time, and $\Lambda_m$ changes little.  An example of this behavior is given in Table \ref{table{sing-values} } in Appendix \ref{sec:stochastic example}.  The product of the three $\Lambda$'s is the Jacobian $J$, which generally changes little because $J(\vec{x}_0,t)=\ln\big(\rho(\vec{x}_0,t)/\rho_0\big)$, where $\rho$ is the density of the fluid at time $t$ and $\rho_0$ is its initial density.  

In realistic systems, the force constraint cannot be satisfied if $B^2$ increases exponentially, which implies the magnetic field line flow must become oriented so $\hat{u}\cdot\vec{B}_0\rightarrow0$.  The magnetic field component proportional to $\hat{s}\cdot\vec{B}_0$ goes to zero, so it is irrelevant, but the field component $\hat{m}\cdot\vec{B}_0$ gives neither a large increase nor decrease in the magnetic field strength.  Indeed, no change is required.  In other words, the magnetic field and the flow of the field lines must adjust such that $\vec{B} = B\hat{M}$.  The cause of this adjustment is no more obscure than the adjustment in the direction of the velocity of water as it flows along a curving channel with steep side walls.  

A 1994 paper by Longcope and Strauss \cite{Longcope-Strauss} did study the effect of chaotic advection on magnetic reconnection, but they used a two-dimensional model, which eliminates the middle singular value and the direction $\hat{M}$.  In fluids, which have no force constraints, a two-dimensional model of chaotic advection is adequate and is the standard model \cite{Aref:2017}.  It is also true that most of the literature and the reviews on magnetic reconnection \cite{Zweibel:review,Loureiro:2016} are two dimensional.  Nevertheless, fundamentally different reconnection behavior is be expected between two and three dimensional systems when $R_m\rightarrow\infty$.


\subsection{Constraint of field-line exponentiation} 

A second rationale for ignoring the implications of the work of Aref and Stern could be that trajectories defined by $\hat{M}(\vec{x})$ at a fixed time, $d\vec{x}/ds = \hat{M}(\vec{x})$ do not exponentiate apart.  This can be checked for specific examples, such as the example given in Appendix \ref{sec:stochastic example}, but it is essentially obvious that neighboring trajectories of $\hat{M}(\vec{x})$ do exponentiate apart when the advection is chaotic.  In a chaotic advection, neighboring points $\vec{x}(\vec{x}_0,t)$ follow exponentially separated trajectories from their locations at $t=0$ to the time $t$ at which the trajectories of $\hat{M}$ are to be calculated.


\subsection{Constraint on advection velocity}

A third rationale for ignoring the implications of the work of Aref and Stern could be that an appropriate chaotic advection velocity cannot exist for certain problems of practical interest, such as in solar corona.  This rationale also fails.  Even simple models, such as $\vec{u}(x,y,z)= u_0\hat{z}+\hat{z}\times\vec{\nabla}\chi(x,y,z)$ can produce chaotic advection.  The region in which stirring takes place can have a limited width, $x^2+y^2<w^2$ and need not have large-scale rotation.   Aref gave interesting examples of this type in his 1984 article \cite{Aref:1984} when his $t$ is interpreted as $z/u_0$ with $u_0$ a constant velocity.  A simple map that also has localization and non-rotating properties is given in Appendix \ref{sec:stochastic example}.  

To be applicable to the solar corona, the advection velocity should have only a localized effect and be non-rotational.  In the corona, the regions in which fast magnetic occurs have dimensions of tens of thousands of kilometers, which is highly localized compared to the size of the sun.  A large rotation of the magnetic field lines around each other is expected to cause kinking and if too strong would lead to turbulence, not the large scale separation of magnetic field lines, which gives the fastest reconnection.

\subsection{Constraint on aspect ratio of reconnecting regions}

A fourth rationale for ignoring the implications of the work of Aref and Stern could be that the time or distance required for chaotic advection to create adequate exponentiation is too long to be important in situations such as the solar corona.  In non-toroidal systems, a separation of scale between the length $L$ and the width $w$ of the structures exhibiting chaotic advection is required, Equation (\ref{L/w}).   

Structures that satisfy the requirement on the $L/w$ ratio are prominent in recent coronal observations \cite{Gou:2019}.   

Force balance is generally lost over an Alfv\'enic time scale $L/V_A$ in a region in which reconnection takes place \cite{Boozer:part.acc.}.  The effects of this loss of equilibrium propagate across the magnetic field lines at the cross-field Alfv\'en speed, which can extend the volume of fast reconnection. 

None of these rationales offer a basis for ignoring the exponentially large change that chaotic advection would produce in the theory of magnetic reconnection.  No knowledgable person denies that neighboring magnetic field lines can exponentiate apart---non-axisymmetric toroidal magnetic-confinement devices are proverbial for the extreme care required in design and construction to avoid this outcome.


\section{X-points  \label{sec:X-points} }

\begin{figure}
\centerline { \includegraphics[width=3.0in]{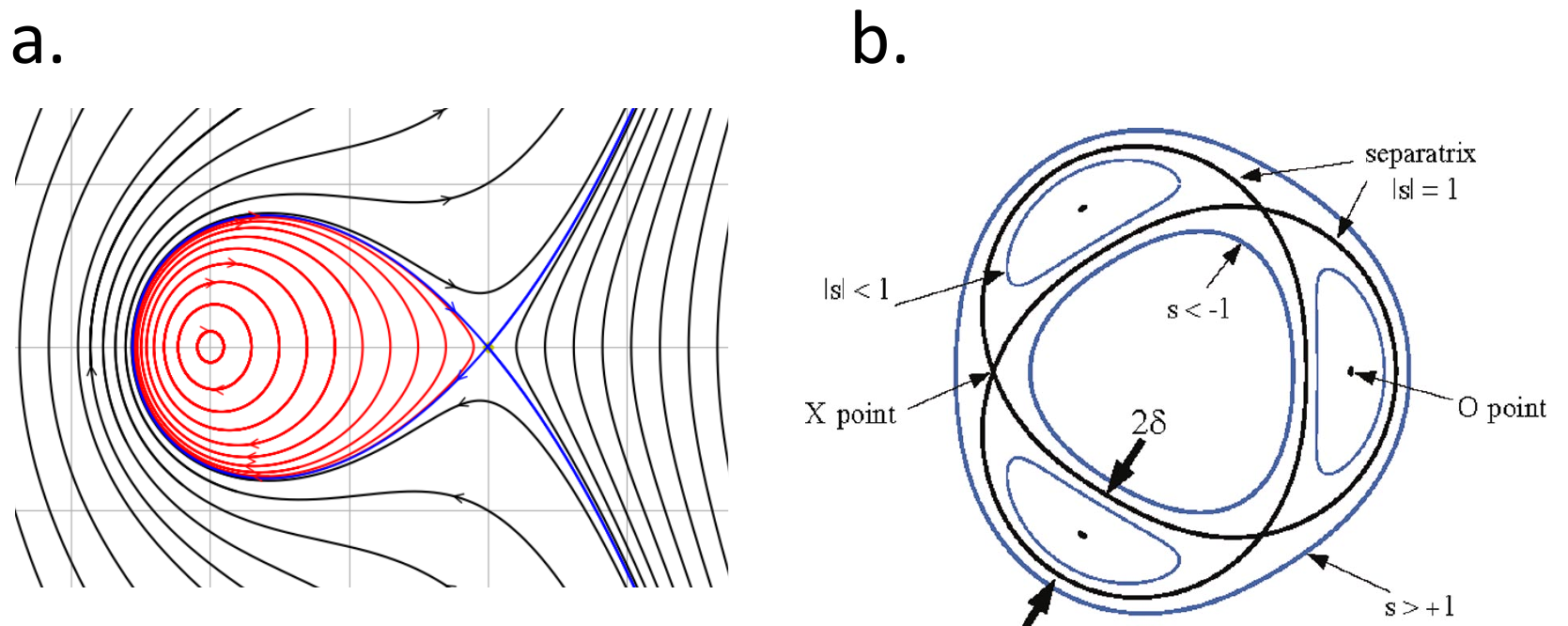} }
\caption{Two types of separatricies are shown.  Figure \ref{fig:separatrix}a illustrates a separatrix and its X-point that separate magnetic field lines which explore fundamentally different regions of space.  Figure \ref{fig:separatrix}b illustrates an island chain in which magnetic field lines both inside and outside the separatricies explore similar regions of space.  Figure \ref{fig:separatrix}b was Figure 2 in  \cite{Boozer:RMP}.  
 }  
\label{fig:separatrix}
\end{figure}

In toroidal plasmas, such as axisymmetric tokamak plasmas, magnetic evolution is peculiar at X-points, Figure (\ref{fig:separatrix}).  An X-point has three properties: (1) The magnetic field line that passes through the X-point closes on itself.  Every point along that field line is an X-point, so an X-point is in reality an X-line. (2) The field lines that pass close to an X-point are of different topological types.  The separatrix which is a curve that connects to an X-point, Figure \ref{fig:separatrix}, encloses magnetic field lines, the interior field lines, that have a fundamentally different topology from those outside the separatrix, the exterior field lines.  (3) The magnetic field lines that pass close to the X-line at one location separate from the X-line exponentially with distance along the line.

Although field lines that pass close to an X-line exponentiate apart, the direct effect on reconnection is small for two reasons:  (1)  When the distance of closest approach of a line to the X-line is $\delta_x$, the maximum number of exponentiations is $\sigma \approx \ln(a/\delta_x)$ where $a$ is the minor radius of the torus, so a volume filled by field lines that under $\sigma$ exponentiations is approximately $e^{-2\sigma}$ of the total volume.   (2) During an evolution, magnetic field lines must be separated by a distance $c/\omega_{pe}$ across the separatrix to be distinguishable.  When the length of an island is $a$, magnetic field lines that lie a distance $c/\omega_{pe}$ on either side of the separatrix have a distance of closest approach to the X point $\delta_x\approx\sqrt{a (c/\omega_{pe})}$.  The entire region around an X-point consists of indistinguishable magnetic field lines, so the details of their individual trajectories cannot be relevant.  

The most important issue relative to an X-line is whether magnetic field lines that have different topological types and pass near the X-line, nevertheless, remain in the same region of space.  Figure \ref{fig:separatrix}a illustrates the case in which the field lines do not, and Figure \ref{fig:separatrix}b illustrates a case in which they do.  When the field lines that come close to an X-line stay in the same region of space by forming an island chain, the evolution is generally on a resistive time scale, the Rutherford rate \cite{Rutherford:island1973}.

In Figure \ref{fig:separatrix}a, field lines that pass near the X-line go to fundamentally different regions of space.  The evolution properties of the interior lines, which are enclosed by the separatrix, can be fundamentally different from those outside of the separatrix, the exterior lines.  When the interior magnetic field lines are embedded in a near ideal plasma, the exterior lines may be unable to carry a steady current because they intercept an insulator or the time required for information about their connections may be far longer than the time scale for the evolution of the magnetic field in the enclosed region.  Information on field line connections propagates along the magnetic field lines at the shear Alfv\'en speed.  

The rapid reconnection that can occur over the region in which field lines come within a distance $\delta_x\approx\sqrt{a (c/\omega_{pe})}$ of an X-point makes it impossible for separatrices to prevent Alfv\'enic reconnection by themselves.  The peculiar role of X-points in the evolution of toroidal plasmas was recognized in 1975 by Grad \cite{Grad:1975}.   Simulations of evolving toroidal plasmas, especially plasmas of very large aspect ratio, could clarify effects associated with differing responses of plasmas interior and exterior to a separatrix including the implications of very long transit times for shear Alfv\'en waves.  Very restricted simulations of this type have been carried out for tokamak plasmas \cite{Pfefferle:2018}, but a comprehensive study has never been a focus.


\section{Null Points \label{sec:null prop} }

In naturally occurring plasmas, the existence of magnetic field lines that close on themselves seems highly unlikely, and that is a requirement for an X-line.  An important question in naturally occurring plasmas is whether null points of a magnetic field, $B^2\equiv B_x^2+B_y^2+B_z^2=0$, have special reconnection properties.  A line along which $B^2=0$ can be removed altogether or broken into point nulls by an arbitrarily small perturbation, so isolated point nulls is the only case of any practical importance.  As will be shown null points have reconnection properties related to X-lines. 

In 1988, John Greene noted \cite{Greene:1988} the close relationship between X-points, which are connected by a separatrix, and nulls, which are connected by separators: ``\emph{the role of the separatrix is played by a line of force that connects two isolated nulls}."   The geometry of separators was discussed in 1990 by Lau and Finn \cite{Lau-Finn:1990}.    The role of separators in reconnection was the subject on a 1993 paper by John Greene \cite{Greene:1993}.  David Pontin \cite{Pontin:2012} and Eric Priest  \cite{Priest:2016} have reviewed the theory of magnetic reconnection in the presence of nulls.  

The focus of the literature on reconnection with nulls is on the formation of a singular current density, $j \propto R_m$, which was thought to be necessary to have a rapidly reconnecting field in the limit $R_m\rightarrow\infty$.  But, the current density need increase only as $\ln R_m$ for reconnection even in the absence of nulls.  The properties of magnetic fields near null points must be treated in a non-conventional manner to make their relation to X-points explicit.  

Magnetic field lines that remain distinguishable in the presence of an evolution pass by nulls at a relatively large distance,  $\delta_n \approx (a^2 c/\omega_{pe})^{1/3}$, Section \ref{sec:nulls-disting}, so the properties of trajectories that pass closer to a null than $\delta_n$ cannot be of direct relevance to the theory of magnetic reconnection.  Although the precise trajectories near a magnetic null are of limited relevance, the trajectories are of interest and are studied in Appendix \ref{near-null traj}.

Magnetic field lines that pass within the vicinity of a null can explore entirely different regions of space.  This is the case for a dipole magnetic field interacting with an externally-produced constant magnetic field, which has been studied by Todd Elder working with Allen Boozer.  A publication is being prepared.  When the dipole moment and the constant field are perfectly aligned, a circular line null is produced about the dipole.  For all other cases, there are a pair of nulls.   The field line properties can be understood by considering four spheres: (1) A sphere of radius $a$ about the dipole, which represents the object in which the dipole is embedded, for example the earth.  (2) The sphere around each of the two nulls, which have a radius $r_0<<a$. (3) A sphere of radius $R>>a$, which records how the magnetic field that interacts with the dipole are mapped to lines that come from or go to infinity.  

A different situation arises if a region of space contains a large number of magnetic field nulls.  The magnetic field lines of even a curl free magnetic field can  place nulls in a volume with a complicated distribution and interact with changes that are driven by changes in the boundary conditions that have not been explored in ways they could be Section \ref{many nulls}.


\subsection{Distinguishable field lines and nulls \label{sec:nulls-disting} }

The conservation of magnetic flux along a magnetic flux tubes shows that magnetic field that are distinguishable during an evolution have a typical closest approach to a null of $\delta_n \approx (a^2 c/\omega_{pe})^{1/3}$.

The magnetic flux that lies within a distance $c/\omega_{pe}$ of a separator is proportional to the spatial scale of the separator $a$ times $c/\omega_{pe}$. 
This flux of indistinguishable magnetic field lines must pass by a null at a typical distance $\delta_n$.  Equating the two fluxes, $(\delta_n/a)\delta_n^2\approx a c/\omega_{pe}$.  The factor $\delta_n/a$ is proportional to the magnetic field strength near the null relative to the field strength along the separator, and the factor $\delta_n^2$ is proportional to the area of a sphere around the null.

Magnetic field lines that pass within a distance $\delta_n \approx (a^2 c/\omega_{pe})^{1/3}$ of a null become indistinguishable in a reconnection, so their precise trajectories cannot be relevant to the theory of reconnection.  The effect of nulls on reconnection, as with X points, is whether the field lines that pass within the vicinity of a null explore entirely different regions of space or not.


\subsection{The electric potential at a null}

The electric potential at the null is a constant $\Phi_0$ and is determined by the condition that no net current enter or leave the null, $\oint  \vec{j}_0\cdot\hat{r} \sin\theta d\theta d\varphi=0$, which is equivalent to $\vec{\nabla}\cdot\vec{j}=0$ there.  The $\vec{\nabla}\cdot\vec{j}=0$ condition holds throughout plasmas of common interest, Section 8.1.3 of \cite{Boozer:NF3D}.


\subsection{Ohm's law}

A general Ohm's law has the form $\vec{E}+\vec{v}\times\vec{B} =\vec{\mathcal{R}}$, where $\vec{v}$ is the plasma velocity \cite{Schindler:1988}.  In non-relativistic theory, which is used in this paper, $\vec{\mathcal{R}}$ is the electric field in a frame that moves with the plasma.   Let $\Phi$ be a solution to $\vec{B}\cdot\vec{\nabla}\Phi=\vec{B}\cdot\vec{\mathcal{R}}-\mathcal{E}_{ni} B$, where $\mathcal{E}_{ni}$ is constant along each magnetic field line.  Define a velocity perpendicular to the magnetic field $\vec{u}_\bot$ by
\begin{eqnarray}
&&\vec{v}_\bot =  \vec{u}_\bot +\vec{B}\times \frac{\vec{\mathcal{R}}+\vec{\nabla}\Phi}{B^2}, \hspace{0.2in}\mbox{then}\hspace{0.2in}\\
&&\vec{E}+\vec{u}_\bot\times\vec{B}=-\vec{\nabla}\Phi +\mathcal{E}_{ni}\vec{\nabla}\ell. \label{E}
\end{eqnarray}
The distance along a magnetic field line is $\ell$, and
\begin{equation}
\mathcal{E}_{ni} \equiv \frac{\int \vec{E}\cdot\frac{\vec{B}}{B}d\ell}{\int d\ell},
\end{equation}
where both integrals are calculated using the same limits of integration.  The integration limits can be (1) $\ell\rightarrow \pm\infty$ as on the irrational magnetic surfaces of a toroidal plasma, (2) a wall on which $\Phi$ has a specified value, such as $\Phi=0$ on a perfectly conducting grounded wall, or (3) the potential $\Phi_0$ on the infinitesimal sphere surrounding a null. 

Only terms involving $\mathcal{E}_{ni}$ produce a non-ideal magnetic evolution. 




\subsection{Magnetic Field with many nulls \label{many nulls} }

A simple model of a magnetic field with many nulls could be studied computationally to gain information on magnetic reconnection in such fields.

Let the initial magnetic field be curl-free and in a perfectly conducting rectangular box $0<x<a$, $0<y<a$, and $0<z<h_0$, where
\begin{eqnarray}
\vec{B}_0 &=& \sum_{mn}  \vec{\nabla}\phi_{mn}(x,y,z); \\
\phi_{mn} &=& \frac{B^{(0)}_{mn}}{k_{mn}} \cos\left(\frac{m\pi}{a} x\right)\cos\left(\frac{n\pi}{a}y\right) \times \nonumber \\
&& \hspace{0.3in} \times \frac{\sinh(k_{mn}(z-h_0/2))}{\cosh(k_{mn}h_0/2)};  \label{phi box} \hspace{0.2in}\\
k_{mn}&=& \frac{\pi}{a} \sqrt{m^2+n^2};
\end{eqnarray}
\begin{eqnarray}
(\hat{z}\cdot\vec{B}_{mn})_{0,h_0} &=&B^{(0)}_{mn} \cos\left(\frac{m\pi}{a} x\right)\cos\left(\frac{n\pi}{a}y\right). \hspace{0.2in}
\end{eqnarray}
There is no normal field on the sides of the box, but there is a normal field on the bottom and top, $(\hat{z}\cdot\vec{B}_{mn})_{0,h_0}$, on the $z=0$ and  $z=h_0$ surfaces.  By adjusting the $B^{(0)}_{mn}$'s, a magnetic field with many nulls can be achieved.   When $m\leq J_{max}$ and $n\leq J_{max}$, then $\delta_J\equiv a/(\pi J_{max})$ gives a natural distance scale between nulls, and most of these nulls lie within a distance of order $\delta_J$ of either the bottom or the top surface.  The number of Fourier coefficients, the $B^{(0)}_{mn}$'s scale as $J_{max}^2$.  A specific model is to make all the $B^{(0)}_{mn}$'s have an equal amplitude.

Reconnection could be studied by assuming the box is filled with a highly conducting plasma with the height of the box, $h(t)$, a function of time while $a$ is held fixed.  Since there is no normal field to the sides of the box, variations in the height do not change the normal field on the side walls.  Since the area of the top and bottom surfaces do not change, the normal field to those surfaces does not change because of the assumption that all surfaces of the box are perfect conductors.

The curl-free magnetic field given by the potential of Equation (\ref{phi box}) gives insights. For various values of $J_{max}^2$ and $h_0/a$, field line integrations allow a number of questions to be addressed.  The most important question addresses the issue of  the extent of the regions in which reconnection would occur due to the indistinguishability of magnetic field lines on the $c/\omega_{pe}$ scale.  To do the study, launch a large number of magnetic field lines in a small square at $z=0$ that has an area $\delta_b^2 <<\delta_J^2$ and has a magnetic field directed into the box.  Record the locations of the exit points, which must have a magnetic field directed out of the box.  

The exit-point data can be used to determine the magnetic flux enclosed  in regions covered by exit points.  With perfect precision the exit-point flux equals the entry-point flux, but if the magnetic field lines are stochastic within the box the magnetic flux in regions covered by exit points can be larger.  By the area covered by exit points is meant the area obtained by counting the small square boxes on the exit surface through which a field line passes as the area of each counting box becomes small.   The number of trajectories being followed must be increased proportionately as the size of the counting boxes becomes small.  

The region subject to rapid reconnection due to the indistinguishability of magnetic field lines on the $c/\omega_{pe}$ scale is larger than the region covered by the exit point flux.  Suppose the size of the squares in which field lines are launched $\delta_b$ equals $c/\omega_{pe}$, then the launched field lines are indistinguishable.  If both the bottom $z=0$ and $z=h_0$ surfaces are covered with squares that are $c/\omega_{pe}$ on a side, then any field line that exit in one of these squares is indistinguishable during an evolution from every other field line in that square.  Consequently, it is sum of the areas of the exit squares at the $c/\omega_{pe}$ scale that determines the area in which evolution will cause a reconnection due to the effect of the $c/\omega_{pe}$ indistinguishability of magnetic field lines.

Launches at more than one small square should be considered, but the typical behavior should quickly become apparent.  Other questions are interesting, though not of such obvious importance. (1) What is the distribution of magnetic field line lengths $L$ between entry and exit from the box?  (2)  What fraction of the lines launched at the bottom, $z=0$, exit at the top, $z=h_0$?  (3) Are there any regions within the box that are not covered by field lines launched from either the top or the bottom?


\section{Summary \label{summary} }

Both X-points and null points represent places where magnetic field lines of fundamentally different topologies and, therefore, evolution properties come arbitrarily close to one another.  Other points  on these field lines can be separated by distances comparable to the overall system size.  

Electron inertia prevents even a collision-free plasma from constraining reconnection on distance scales smaller that $c/\omega_{pe}$, which implies field lines with fundamentally different topologies can not be prevented from reconnecting on that scale independent of the ideality of the plasma in any other sense.  This non-locality implies that magnetic field lines that come within a distance $\delta_x\approx\sqrt{a (c/\omega_{pe})}$ of an X-point or  $\delta_n \approx (a^2 c/\omega_{pe})^{1/3}$ of a null cannot directly affect magnetic reconnection, where $a$ is a typical system dimension.

The traditional assumption that reconnection requires the dissipation or diffusion of the reconnected magnetic flux implies that reconnection can be competitive with an ideal evolution only when the maximum current density is proportional the magnetic Reynolds number $R_m$.   In three dimensions, magnetic flux can be spatially mixed in an ideal evolution \cite{Boozer:part.acc.} by the exponential separation of neighboring magnetic field lines \cite{Boozer:ideal-ev}.  The implication is that reconnection  becomes competitive when the current density is proportional to $\ln R_m$.  Flux mixing conserves magnetic helicity \cite{Boozer:part.acc.}, so helicity conservation is intrinsic as $R_m\rightarrow\infty$.  

Magnetic field lines that pass in the vicinity of a null or an X-point can reach fundamentally different regions of space.  Depending on the assumptions about these spatial regions, null and X-points may not be able to impede reconnection over a longer time than inertial, which means a time scale set by the shear Alfv\'en wave.


\vspace{0.2in}

\section*{Acknowledgements}

This material is based upon work supported by the U.S. Department of Energy, Office of Science, Office of Fusion Energy Sciences under Award Numbers DE-FG02-95ER54333, DE-FG02-03ER54696, DE-SC0018424, and DE-SC0019479.

\appendix

\section{Traditional reconnection theories \label{sec:traditional-reconnection} }

Traditional reconnection theory, which dominates the literature and the reviews, assumes the current density associated with the reconnection is concentrated in a narrow channel and has a characteristic magnitude $j\approx (B/\mu_0L)R_m$.  The current density must be this intense to resistively diffuse the reconnecting flux $BL^2$ by the resistive voltage $\eta jL$ at a rate that competes with the evolution time $\tau_{ev}$.   

Two issues should be considered.  (1) How does a natural evolution produce such a large current density.  Traditional theory often assumes the large current density just exists, as in a Harris sheet, and does not explain how this arises in a natural evolution.  (2) The width of the resistive layer produced in an evolution time $\delta_\eta\equiv \sqrt{(\eta/\mu_0)\tau_{ev}}=L/\sqrt{R_m}$, and  $j_{ex}\equiv (B/\mu_0\delta_\eta)=(B/\mu_0L)\sqrt{R_m}$ would be the expected current density, which is a factor of $\sqrt{R_m}$ smaller than the current density required for resistivity to compete with evolution.

The resolution of this paradox led to plasmoid theory, which focuses on obtaining a large gradient in the flow velocity of the magnetic field lines $\vec{u}_\bot$ at the location of the current sheet, which has a thickness $\delta_\ell\approx L/R_m$.  A heuristic treatment of plasmoid theory is given in \cite{Liu:2017} and the theory is also discussed in major reviews \cite{Zweibel:review,Loureiro:2016}.

  A magnetic field is required in the direction of flow of the sheet current.  Otherwise the magnetic field would have a line zero,   The local evolution time $\tau_\ell =1/(u_\bot/\delta_\ell)$.  The resistive layer thickness is $\delta_\ell^2 = (\eta/\mu_0)\tau_\ell$, or $\delta_\ell = (\eta/\mu_0)/u_\bot$.  The flow of the magnetic field lines into the resistive layer of thickness $\delta_\ell$ occurs over a scale $L$, so the time scale for the large scale evolution of the magnetic field is $\tau_{ev}=1/(u_\bot/L)$.  Consequently, $L/\delta_\ell \approx (\mu_0/\eta)u_\bot L \approx (L^2\mu_0/\eta)(u_\bot/L)=R_m$.



\section{Lagrangian coordinates \label{sec:Lagrangian} }

Lagrangian coordinates are the central concept for understanding why stirring creates conditions for enhanced mixing and why the field line velocity  $\vec{u}_\bot$ of an ideal evolution, Equation (\ref{ideal-B}), creates the conditions for an enhanced reconnection.  Let $\vec{x} =x\hat{x}+y\hat{y}+z\hat{z}$ denote the three Cartesian coordinates.  Lagrangian coordinates are defined by an integration along the streamlines
\begin{equation}
\frac{\partial\vec{x}(\vec{x}_0,t)}{\partial t} \equiv \vec{u}(\vec{x},t) \mbox{     where    } \vec{x}(\vec{x}_0,t=0)=\vec{x}_0.
\end{equation}
The three-by-three Jacobian matrix of Lagrangian coordinates can be decomposed as
\begin{eqnarray}
\frac{\partial \vec{x}}{\partial \vec{x}_0} &\equiv& \left(\begin{array}{ccc}\frac{\partial x}{\partial x_0} & \frac{\partial x}{\partial y_0} & \frac{\partial x}{\partial z_0} \vspace{0.03in}  \\ \vspace{0.03in} \frac{\partial y}{\partial x_0} & \frac{\partial y}{\partial y_0} & \frac{\partial y}{\partial z_0}  \\ \frac{\partial z}{\partial x_0} & \frac{\partial z}{\partial y_0} & \frac{\partial z}{\partial z_0}\end{array}\right) \nonumber\\
  &=& \tensor{U}\cdot\left(\begin{array}{ccc}\Lambda_u & 0 & 0 \\0 & \Lambda_m & 0 \\0 & 0 & \Lambda_s\end{array}\right)\cdot\tensor{V}^\dag .  \\
  &=&\hat{U}\Lambda_u \hat{u} + \hat{M}\Lambda_m \hat{m} + \hat{S}\Lambda_s \hat{s} \label{SVD.of.Jacobian}
\end{eqnarray}
where $\tensor{U}$ and $\tensor{V}$ are unitary matrices, $\tensor{U}\cdot\tensor{U}^\dag=\tensor{1}$.   The three real coefficients $\Lambda_u \geq \Lambda_m \geq \Lambda_s \geq 0$ are the singular values of a Singular Value Decomposition (SVD).  The orthogonal unit vectors associated with the unitary matrix $\tensor{U}$ are $\hat{U}$, $\hat{M}$, and $\hat{S}$.  The orthogonal unit vectors associated with the unitary matrix $\tensor{V}$ are $\hat{u}$, $\hat{m}$, and $\hat{s}$.  In a chaotic flow, $\Lambda_u$ increases exponentially in time, $\Lambda_s$ decreases exponentially in time, and $\Lambda_m$ is relatively slowly changing.  For an example, see Table \ref{table{sing-values} } in Appendix \ref{sec:stochastic example}.

In a pure advection, $\partial n/\partial t=- \vec{\nabla}\cdot(n\vec{u})$, the density obeys $n(\vec{x},t) = n_0(\vec{x}_0)/J$, where $n_0(\vec{x}_0)$ is the initial density profile, $J$ is the Jacobian (determinant) of the Jacobian matrix, which means $d^3x = J d^3x_0$.  The gradient of the density steepens exponentially $\vec{\nabla}(Jn) = (\partial \vec{x}_0/\partial \vec{x})\cdot \partial n_0/\partial \vec{x}_0\rightarrow \hat{S}(\hat{s}\cdot\partial n_0/\partial \vec{x}_0)/\Lambda_s$ as $\Lambda_s\rightarrow0$.  Chaotic advection leads to exponentially steepening gradients, which makes diffusion exponentially faster.


\section{Reconnection on $c/\omega_{pe}$ scale \label{c/omega-p} }

Equation (\ref{collisonless ev}) implies that a magnetic evolution is spread by a distance $c/\omega_{pe}$ across the direction in which the evolution takes place.   This will be proven assuming that $c/\omega_{pe}$ is very small compared to the gradient scale of the magnetic field.

The ideal part of the magnetic evolution is given by
\begin{equation}
\frac{\partial \vec{\mathcal{B}} }{\partial t} \equiv \vec{\nabla}\times (\vec{v} \times\vec{B}),  \label{mathcal(B) ev}
\end{equation} 
which has components along and perpendicular to the magnetic field $\vec{B}$.  For reconnection, the important issue is the change in the magnetic field perpendicular to itself.  Let $(x,y,z)$ be local coordinates, where $z$ is along $\vec{B}$, $x$ is in the direction given by the part of $\partial \vec{\mathcal{B}}/\partial t$ that is perpendicular to $\vec{B}$, and $y$ is the other coordinate perpendicular to $\vec{B}$.  One can then write,
\begin{eqnarray}
\vec{B}(x,y,z,t) &=& \int G(x - \xi) \vec{\mathcal{B}}(\xi,y,z,t)d\xi.
\end{eqnarray}
Equations (\ref{collisonless ev}) and  (\ref{mathcal(B) ev}) then imply
\begin{eqnarray}
\frac{\partial \vec{\mathcal{B}} }{\partial t} &=& \frac{\partial \vec{B} }{\partial t} + \left( \frac{c}{\omega_{pe}}\right)^2 \vec{\nabla}\times\left(\vec{\nabla}\times\frac{\partial \vec{B}}{\partial t}\right)  \\
&=&\int \Big\{G  \frac{\partial \vec{\mathcal{B}}}{\partial t} - \frac{\partial \vec{\mathcal{B}}}{\partial t} \nabla^2G \nonumber \\
&& \hspace{0.4in}  +\frac{\partial \vec{\mathcal{B}}}{\partial t}\cdot\vec{\nabla} (\vec{\nabla}G)\Big\}d\xi.  \label{Perp-direction}
\end{eqnarray}
The coordinate $x$ was chosen so the last term in Equation (\ref{Perp-direction}) is zero, and
\begin{eqnarray}
&&\frac{\partial^2 G}{\partial x^2}- G = \delta(x - \xi),  \mbox{    so   } \\
&& G(x - \xi) = \frac{1}{2(c/\omega_{pe})} \exp\Big( - \frac{| x - \xi |}{c/\omega_{pe}} \Big), \mbox{    using   }\hspace{0.3in}\\
&& \Big[ \frac{\partial G}{\partial x} \Big]_{x=\xi} =1,
\end{eqnarray}
which follows from integrating across the Dirac delta function.   That is,
\begin{eqnarray}
\frac{\partial\vec{B}(x,y,z,t)}{\partial t} &=& \int \frac{\exp\Big( - \frac{| x - \xi |}{c/\omega_{pe}} \Big)}{2(c/\omega_{pe})} \frac{\partial \vec{\mathcal{B}}(\xi,y,t) }{\partial t} 
d\xi.\nonumber\\
\end{eqnarray} 
The magnetic evolution is spread  by a distance $c/\omega_{pe}$ across the magnetic field lines at each point along the magnetic field line, which implies that if two lines come within a distance of $c/\omega_{pe}$ of each other anywhere along their trajectories, then the evolution reconnects them.


\section{Required force on the boundary \label{Bndry-Force} }

The force required to drive a magnetic field to a state in which the magnetic field has sufficient exponentiation for a fast magnetic reconnection is relatively small $\approx \ln(R_m)(B^2/\mu_0) w/L$ per unit area, where $B$ is the strength of the magnetic field $w$ is the width of the region across which the field lines exponentiate and $L$ is length of the lines in the region.  To be specific, the region will be called the solar corona though the theory has a wider applicability.

What will be shown is that one can surround a coronal region with a fixed surface, called a control surface, that is perfectly conducting.  The changing effects of currents outside the control surface can be represented by a flow in that surface.  The location of the control surface can be chosen so a magnetic field line that leaves and returns to the solar photosphere has the two photospheric foot points on the control surface and remains within the region enclosed by the control surface as it goes between the two footpoints.   

Suppose a set of field lines leave the control surface in a small spot where the control surface is not moving but they enter the surface in a region of width $w$, which is moving with a chaotic velocity.  This velocity separates neighboring points on the control surface exponentially in time but the points can never be separated by a distance greater than $w$.  If the plasma is perfectly conducting, no reconnection is possible.   After a sufficient evolution time, field lines leaving the stationary region in the arbitrarily small spot can reach a separation $w$.  That is, the exponentiation can cause the separation to reach arbitrarily large values. 

As discussed in the Introduction, the separation $\vec{\delta}$ between two lines obeys the equation $ B d \vec{\delta}/d\ell =( \vec{\delta}\cdot\vec{\nabla})\vec{B}$, where $\ell$ is the distance along a magnetic field line.  The magnitude of tensor $\vec{\nabla}\vec{B}$ is approximately $\mu_0 j_{||}$, which implies $\ln(w/\delta_0) \approx (\mu_0j_{||}/B)L$.  The magnetic field produced by the current is $B_c \approx \mu_0 j_{||}/w$ so $B_c= \ln(w/\delta_0) (w/L)B$.  It is the field $B_c$ that should be used to estimate $\Big[\vec{B}_t\Big]$ in Equation (\ref{Force}) for the force.


\subsection{Control surface} 

The fluid mechanics concept of a control surface $\vec{x}_s(\theta,\varphi)$ can be used to understand how distant currents interact with reconnecting magnetic fields within the volume enclosed by that surface, the control volume.  The shape of the control surface is chosen to clarify the physics. For example in the coronal problem, the control surface can run in part along the photospheric surface and in part along a surface above the corona.

\subsubsection{Coordinates near the control surface}

Positions in the vicinity of the control surface can defined using 
\begin{eqnarray}
\vec{X}_s(\xi,\theta,\varphi) \equiv \vec{x}_s(\theta,\varphi) + (\xi-a) \frac{\frac{\partial \vec{x}_s}{\partial\theta}\times\frac{\partial \vec{x}_s}{\partial\varphi}}{\left(\frac{\partial \vec{x}_s}{\partial\theta}\times\frac{\partial \vec{x}_s}{\partial\varphi}\right)^2}.
\end{eqnarray}
The Jacobian of these coordinates within the surface $\xi=a$ is then
\begin{eqnarray}
&&\frac{\partial \vec{X}_s}{\partial\xi}\cdot\left(\frac{\partial \vec{X}_s}{\partial\theta}\times\frac{\partial \vec{X}_s}{\partial\varphi}\right)=1  \hspace{0.1in}\mbox{and}\hspace{0.1in} \\
&&\vec{\nabla}\xi = \frac{\frac{\partial \vec{X}_s}{\partial\theta}\times\frac{\partial \vec{X}_s}{\partial\varphi}}{\frac{\partial \vec{X}_s}{\partial\xi}\cdot\left(\frac{\partial \vec{X}_s}{\partial\theta}\times\frac{\partial \vec{X}_s}{\partial\varphi}\right)} =\frac{\partial \vec{x}_s}{\partial\theta}\times\frac{\partial \vec{x}_s}{\partial\varphi} \hspace{0.2in} \label{xi-orthog}
\end{eqnarray}
is the normal to the $\xi=a$ surface.  The dual relations of general coordinates, which are derived in the Appendix to \cite{Boozer:RMP}, were used.  These relations also imply that within the $\xi=a$ surface
\begin{eqnarray}
\frac{\partial \vec{X}_s}{\partial\xi} &=&\vec{\nabla}\xi, \\
 (\vec{\nabla}\xi)\cdot(\vec{\nabla}\theta)&=&0, \hspace{0.1in}\mbox{and}\hspace{0.1in} (\vec{\nabla}\xi)\cdot(\vec{\nabla}\varphi)=0.
\end{eqnarray}
These coordinates in general have singularities.  For a spherical surface, $(\partial \vec{x}_s/\partial\theta)\times(\partial \vec{x}_s/\partial\varphi)=r^2 \sin\theta \hat{r}$.  But, these singularities will not cause problems in our analysis; they can be assumed to be located at a place where the control surface has no direct effect on the reconnection.

\subsubsection{Control surface as a perfect conductor}

Analyses using a control surface are most straight forward when that surface can be taken to be a perfect conductor.  The difficulty of doing this comes from the magnetic field normal to the control surface.  

When the control surface is not a perfect conductor, the normal field can  be controlled by the addition of a divergence-free normal magnetic field, $\vec{B}_n=B_n(\theta,\varphi,t)\vec{\nabla}\theta\times\vec{\nabla}\varphi$ with $\oint B_nd\theta d\varphi=0$.   But, this additional field complicates the plasma analysis and will be assumed to be unnecessary at least in the regions of the control surface, impacted by the magnetic field lines of interest. 

  The closed surface can be taken to be a perfectly-conducting boundary when a flow velocity $\vec{u}_s$ within the surface, $\vec{u}_s\cdot\vec{\nabla}\xi=0$,  can be found that reproduces any normal and tangential field on the plasma side of that surface with $\partial \vec{B}/\partial t=\vec{\nabla}\times(\vec{u}_s\times\vec{B})$. 
  
The primary interest will be in the tangential magnetic field to the control surface; the primary complications are in the normal component of the magnetic field, $\vec{B}\cdot\vec{\nabla}\xi$, which can be controlled at least locally by flow in the control surface.  The flow $\vec{u}_s$ is within the surface, the surface itself, $\vec{x}_s(\theta,\varphi)$, is time independent, so
\begin{eqnarray}
\frac{\partial \vec{B}\cdot\vec{\nabla}\xi}{\partial t} &=&\vec{\nabla}\xi\cdot \vec{\nabla}\times(\vec{u}_s\times\vec{B}) \nonumber\\
&=& - \vec{u}_s\cdot\vec{\nabla}(\vec{B}\cdot\vec{\nabla}\xi)- (\vec{B}\cdot\vec{\nabla}\xi) \vec{\nabla}\cdot\vec{u}_s.\hspace{0.3in}
\end{eqnarray}
The flow in the surface is written as the sum of two parts, one curl free and one divergence free,
\begin{eqnarray}
&& \vec{u}_s = \vec{\nabla}\chi(\theta,\varphi,t) + \vec{\nabla}\xi\times\vec{\nabla}\phi(\theta,\varphi,t), \hspace{0.1in}\mbox{where}\hspace{0.4in} \\
&& \nabla^2\chi = \vec{\nabla}\cdot \vec{u}_s.
\end{eqnarray}
Equation (\ref{xi-orthog}) ensures that $\vec{u}_s\cdot\vec{\nabla}\xi=0$.  

Since it is the tangential field to the control surface that is of primary interest, it will be assumed that the control surface is a perfect conductor and 
\begin{eqnarray}
\vec{u}_s &=&\vec{\nabla}\xi\times\vec{\nabla}\phi(\theta,\varphi,t).
\end{eqnarray}

\subsection{Force in the control surface}

A force must be supplied in the control surface to drive the evolution of the magnetic field.  The magnitude of this force can be easily calculated.

The force per unit volume exerted by a magnetic field is given by the divergence of the magnetic stress tensor $\vec{\nabla}\cdot\tensor{T}$, where 
\begin{equation}
\tensor{T}=\frac{\vec{B}\vec{B}}{\mu_0} - \frac{B^2}{2\mu_0}\tensor{1}
\end{equation}

The force per unit area exerted tangentially to the control surface, which is a shell of infinitesimal thickness, is
\begin{eqnarray}
\vec{\mathcal{F}} &=& \Big[ \tensor{T}\cdot\hat{n}\Big]\\
&=& \frac{B_\bot}{\mu_0} \Big[\vec{B}_t\Big],  \label{Force}
\end{eqnarray} 
where  $\Big[\vec{B}_t\Big]$ is the jump in the tangential magnetic field from one side of the control surface to the other.  The jump in the normal field $\vec{B}_\bot$ must be zero to satisfy $\vec{\nabla}\cdot\vec{B}=0$.  The unit normal $\hat{n}=\vec{\nabla}\xi / |\vec{\nabla}\xi |$.


\section{Stochastic map \label{sec:stochastic example} }

\begin{table}
  \centering 
  \begin{eqnarray}
\left(\begin{array}{cccc} x_0 & 0.1 & 0.5 & 2.0 \\ \Lambda_u & 2.824\times 10^3 & 22.65 & 404.6 \\ \Lambda_m & 1.104& 1.216 & 1.177 \\ \Lambda_s & 3.375\times10^{-4} & 0.03631 & 0.002099 \\ \hat{M}_x & -0.8830 & 0.4312 & 0.5428 \\ \hat{M}_y & -0.1996 & -0.5162 & 0.4432 \\ \hat{M}_z & -0.4247 & -0.7400 & 0.7134\end{array}\right) \nonumber
\end{eqnarray}
  \caption{The three Lyapunov exponents (unstable, medium, and small) and the components of the vector $\hat{M}$ are given for the map defined by Equation (\ref{simple-map}) with $K=2.5$ at different value of $x_0$ but with $y_0=0$.  The Jacobian matrix was calculated after ten iterations using a change in $x_0$ and in $y_0$ of $1.6\times10^{-11}$, which reliably gave the Jacobian of the Jacobian matrix as unity.   }\label{table{sing-values} }
\end{table}


Aref \cite{Aref:1984} gave continuous Hamiltonians that have trajectories that remain in a limited region of space and are of a non-encircling type.  Maps allow much simpler exploration of magnetic field line properties, and a map in which the trajectories remain in a limited region of space and are non-encircling will be given.

The map equations that will be derived  represent magnetic field lines that are carried along by a plasma flow in $x-y$ planes with different planes having different flows.  The planes are separated by a fixed distance in $h$ in the $z$ direction, so the iteration number of the map $n$ is related to $z$ by $z=z_0+nh$.  The essential feature of any map equations for magnetic field lines is that they have a unit Jacobian between the iterates $x_{n+1}(x_n,y_n)$, $y_{n+1}(x_n,y_n)$.

Typical examples of stochastic trajectories have an unperturbed case in which trajectories encircle a line.  Though such examples are very important for understanding stochasticity in toroidal plasmas, they are probably irrelevant for understanding the corona.  A line can be encircled tens of times  before a large exponentiation occurs and such magnetic structures would be highly unstable to kinks in the corona.  Typical examples also have trajectories that reach arbitrarily large values of $x$ and $y$ rather than remaining in compact structures.  Structures far smaller than the radius of the sun are observed in the corona.

The example given here has trajectories that just oscillate between two points when the system is unperturbed, $K=0$, but even when $K$ is sufficiently large to produce strong stochasticity, for example $K=2.5$, the region of stochastic field lines is bounded to moderate values of $x$ and $y$.

The overall effect on the magnetic field of the field-line velocity represented by the map is given by the Jacobian matrix of the transformation $\vec{x}_n(x_0,y_0,z_0)=x_n(x_0,y_0,z_0)\hat{x}+y_n(x_0,y_0,z_0)\hat{y}+z_n(z_0)\hat{z}$, where $z_n=z_0 + nh$ as the iteration number $n$ becomes large.  When the trajectories are stochastic, the Jacobian matrix will have three singular values: one will become exponentially large, one will become exponentially small, and one will not exponentiate for $n>>1$.  The product of the three singular values is unity.  The magnetic field will have to point in the $\hat{M}$ direction, which is defined by the non-exponentiating singular value, for both the forces and the magnetic field to remain finite.  


Map equations can have two free functions, $f(x)$ and $g(y)$, when the map is of the form
\begin{eqnarray}
x_{n+1} &=& - y_n - f(x_n + g(y_n)) \\
y_{n+1} &=& x_n + y_n + g(y_n)+ f(x_n+g(y_n)).
\end{eqnarray}
The Jacobian of this map is unity: $\partial x_{n+1}/\partial x_n= - f'$, $\partial x_{n+1}/\partial y_n= -1 - f'g'$, $\partial y_{n+1}/\partial x_n= 1 +f'$, and $\partial y_{n+1}/\partial y_n=1 + g' + f'g'$.

When $f$ and $g$ are zero, this map takes trajectory points around a closed curve every six iterations.  This map can be followed by the reverse of this map every other iteration, so when $f$ and $g$ are zero, the trajectory just oscillates between two points.  The reverse map is
\begin{eqnarray}
x_{n+1} &=&  x_n + y_n + f(x_n)+ g(y_n+f(x_n))\\
y_{n+1} &=&  - x_n - g(y_n + f(x_n)).
\end{eqnarray}

The combination of these two maps is oscillatory when $f$ and $g$ are zero,  but when coupled with the bounded free functions of the same form,
\begin{equation}
f(x) \equiv Kx e^{-x^2} \mbox{   and   } g(y)  \equiv K y e^{-y^2},  \label{simple-map}
\end{equation}
the map  has a complicated region of exponentiating separating trajectories in the $(x_0,y_0)$ plane for $K=2.5$.  A few examples are given in Table \ref{table{sing-values} }.

Derivatives with respect to $z$ can be taken by assuming that between the iterates $n$ and $n+1$ the map equations are $x =x_n + \Delta_x (z-z_n)$ and  $y =y_n + \Delta_y (z-z_n)$.  The quantities $\Delta_x$ and $\Delta_y$ are constants for $z_{n+1}>z>z_n$ and have the values $\Delta_x=(x_{n+1}-x_n)/h$ and $\Delta_y=(y_{n+1}-y_n)/h$.


\section{Trajectories near a null \label{near-null traj} }

Although magnetic field lines that pass within a distance $\delta_n$ of a null do not preserve their identities during an evolution of even a collisionless plasma, their trajectories are still of interest. 


\subsection{ $\vec{B}$ near a null}

The magnetic field sufficiently close to a magnetic null can be written in a universal form, $\vec{B}=(B_c/2a)\tensor{\mathcal{M}}\cdot\vec{x}$ as $|\vec{x}|\rightarrow 0$ where $\vec{x}=0$ is the location of the null and $B_c/2a$ is a constant.

In deriving the universal form for $\tensor{\mathcal{M}}$, it is useful to write $\vec{B}=\vec{\nabla}\phi + \frac{1}{2} \vec{j}_0\times\vec{x}$, where $\nabla^2\phi=0$ and $\vec{j}_0$ is a constant, the current density at the null.  Note $\vec{\nabla}\times(\frac{1}{2} \vec{j}_0\times\vec{x})=\mu_0\vec{j}_0$ and $\vec{\nabla}\cdot(\frac{1}{2} \vec{j}_0\times\vec{x})=0$.

The scalar potential of the magnetic field can be written
\begin{eqnarray}
\phi &=& \left(\begin{array}{c}x' \\y' \\z'\end{array}\right)^\dag \cdot\left(\begin{array}{ccc}\phi_{11} & \phi_{12} & \phi_{13} \\\phi_{12}& \phi_{22} & \phi_{23} \\\phi_{13} & \phi_{23} &\phi_{33}\end{array}\right)\cdot\left(\begin{array}{c}x' \\y' \\z'\end{array}\right)  \\
&=&  \left(\begin{array}{c}x' \\y' \\z"\end{array}\right)^\dag \cdot \tensor{V}\cdot\left(\begin{array}{ccc}\lambda_1 & 0 &\\ 0& \lambda_2 & 0 \\ 0 & 0 &\lambda_3\end{array}\right)\cdot\tensor{V}^\dag\cdot\left(\begin{array}{c}x' \\y' \\z"\end{array}\right) \hspace{0.2in},
\end{eqnarray}
where $\tensor{V}$ is a real orthogonal matrix, $\tensor{V}^\dag\cdot\tensor{V}=\tensor{1}$, which is equivalent to a rotation through three angles. The Cartesian coordinates in which the matrix for $\phi$ is diagonal are
\begin{equation}
\left(\begin{array}{c}x \\y \\z\end{array}\right) \equiv \tensor{V}^\dag\cdot\left(\begin{array}{c}x' \\y' \\z'\end{array}\right).
\end{equation}
The condition that $\nabla^2\phi=0$ is equivalent to $\lambda_1+\lambda_2+\lambda_3=0$.  Consequently, a magnetic field sufficiently near a null point has the general representation  
\begin{eqnarray}
\vec{B} &=& \frac{B_c}{2a} (1+Q)x\hat{x} + \frac{B_c}{2a} (1-Q)y\hat{y}-  \frac{B_c}{a} z \hat{z} \nonumber \\
&& + \frac{\mu_0}{2}  \vec{j}_0 \times \vec{x}, \mbox{   or   } \label{near-null-field}\\
&=& \frac{B_c}{2a} \tensor{M}\cdot\vec{x},  \mbox{    where   }\\ 
\vspace{0.3in}
\tensor{M} &=& \left(\begin{array}{ccc}1+Q & -\frac{a\mu_0}{B_c}j_{0z} & +\frac{a\mu_0}{B_c}j_{0y} \\ +\frac{a\mu_0}{B_c}j_{0z} & 1-Q & -\frac{a\mu_0}{B_c}j_{0x} \\-\frac{a\mu_0}{B_c}j_{0y} & +\frac{a\mu_0}{B_c}j_{0x} & -2\end{array}\right). \label{near-null matrix}
\end{eqnarray}

The best known representation of a magnetic field near a null is the one obtained by Parnell et al \cite{Parnell:1996}, but their representation has the oddity of mixing the current density at the null into the symmetric as well as into the anti-symmetric part of the matrix $\tensor{M}$. 


\subsection{Magnetic field lines}

Magnetic field lines that enter the null sphere at particular values of $\theta$ and $\varphi$ exit the null sphere at particular values of $\theta$ and $\varphi$.  

There is a caveat for magnetic field lines that exactly strike the null.  This issue can be understood in the simple case in which the near-null magnetic field, Equation (\ref{near-null-field}), has $Q=0$ and $\vec{j}_0=0$.  The magnetic field lines that enter the null sphere in a circle of radius $\delta_p<<r_0$ about one of the poles of the null sphere exit the null sphere in a ribbon of width $\delta_{eq}$ about the equator in the same hemisphere in which they entered.  Flux conservation along magnetic flux tubes implies $\delta_{eq}=\delta_p^2/r_0<<\delta_p$.  Equation (\ref{null-traj}) implies $\delta_p=\sqrt{(2/3\sqrt{3})r_c^3/r_0}$ where $r_c$ is the closest approach distance to the null.  No matter how small $r_c$ may be one can resolve the trajectories; only $r_c$ exactly zero can be an issue.    

For simplicity, the entry and exit points of field lines on the null sphere will be discussed for the case in which the current at the null is zero, $\vec{j}_0=0$.  The $\vec{j}_0\neq0$ case can be treated similarly by finding the eigenvalues and eigenvectors of a matrix that is non-symmetric.  The magnetic field near a null is described by the matrix given in Equation (\ref{near-null matrix}) with $\vec{B} = (B_c/2a) \tensor{M}\cdot\vec{x}$.  The elements of $\tensor{M}$ are constants, which means independent of the location in the sphere.  When $\tensor{M}$ is diagonalized, $\tensor{M} = \tensor{P}\cdot\tensor{m}\cdot\tensor{P}^{-1}$ with $\tensor{m}$ diagonal, new coordinates are obtained, which are related to the Cartesian coordinates by spatial constants,
\begin{equation}
\left(\begin{array}{c}\xi \\\eta \\\zeta\end{array}\right) = \tensor{P}^{-1}\cdot\left(\begin{array}{c}x \\y \\z\end{array}\right).
\end{equation}
The equation for magnetic field lines $d\vec{x}/d\tau=\tensor{M}\cdot\vec{x}$ can be solved as $\xi=\xi_0 \exp(-m_1 \tau)$, $\eta=\eta_0 \exp(-m_2 \tau)$, and $\zeta=\zeta_0 \exp(-m_3 \tau)$.  As will the illustrated for the $\vec{j}_0=0$ case, these equations can be solved for two constants of the motion which give the relation between the entry and exit $\theta$ and $\varphi$ on the null sphere.

When the current density at a null is zero, the near-null magnetic field, Equation (\ref{near-null matrix}), can be represented as a diagonal matrix,
\begin{eqnarray}
\tensor{M} &=& \left(\begin{array}{ccc}1+Q & 0 & 0 \\0 & 1-Q & 0 \\0 & 0 & -2\end{array}\right).
\end{eqnarray}
The trajectories in Cartesian coordinates are $x=x_0 \exp((1+Q)\tau)$, $y=y_0 \exp((1-Q)\tau)$, and $z=z_0 \exp(-2\tau)$.  These three equations give two $\tau$ independent quantities, which means they are constants of the motion, $xyz^2$ and $y^{(1+Q)}/x^{(1-Q)}$.  Written in spherical coordinates, the first constant of the motion is
\begin{eqnarray}
C_1 &=& r^3 \sin(2\varphi) \cos\theta \sin^2\theta.
\end{eqnarray}
The second constant of the motion, $y^{(1+Q)}/x^{(1-Q)}=(\frac{1}{2}r^2 \sin(2\varphi)\sin^2\theta)^Q \tan\varphi$, can be simplified using the first,
\begin{equation}
C_2 = \big| r\cos\theta \big|^Q \tan\varphi.
\end{equation}
The explicit $\varphi$ dependence can be eliminated from the first constant of the motion using the second and the identity $\sin(2\varphi)=2\tan\varphi/(1+\tan^2\varphi)$.

The simplest case is $Q=0$.  The second constant of the motion implies $\varphi$ is constant, so the first constant of the motion can then be written as $C'_1=r^3  \cos\theta \sin^2\theta$.  The function $| \cos\theta \sin^2\theta| \leq 2/3\sqrt{3} \approx0.3849$.  Its maximum and minimum occur as $\cos\theta=\pm 1/\sqrt{3}$.  The radial magnetic field $\vec{B}\cdot\hat{r} = (B_c/2a) r (\sin^2\theta-2\cos^2\theta)$ vanishes at $\cos\theta=\pm 1/\sqrt{3}$.  A magnetic field line that enters the null sphere always does so closer to one of the poles than $\cos\theta=\pm 1/\sqrt{3}$ and always exits between latitude $\cos\theta=\pm 1/\sqrt{3}$ and the equator.  In spherical coordinates the trajectories inside the null sphere $r<r_0$ obey
\begin{equation}
r^3 \cos\theta \sin^2\theta = \pm \frac{2}{3\sqrt{3}} r_0^3, \label{null-traj}
\end{equation}
where $r_0$ is the radius of the null sphere.  In the northern hemisphere pairs of values $\theta$ solve $\cos\theta \sin^2\theta = 2/3\sqrt{3}$; each pair is the entry and exit points of a trajectory.  Similarly in the southern hemisphere for $\cos\theta \sin^2\theta = -2/3\sqrt{3}$.


\end{document}